\documentclass[aps,pre,twocolumn,nofootinbib,floatfix]{revtex4}
\usepackage{epsfig}

\begin{document}

\title{Probability distribution of returns in the Heston model with
stochastic volatility}

\author{Adrian A.\ Dr\u{a}gulescu} 
  \altaffiliation{Now at the Constellation Energy Group, Baltimore}
  \email{adrian.dragulescu@constellation.com}
\author{Victor M.\ Yakovenko}
  \homepage{http://www2.physics.umd.edu/~yakovenk}
  \email{yakovenk@physics.umd.edu}
\affiliation{Department of Physics, University of Maryland, 
        College Park, MD 20742-4111, USA}

\date{{\bf cond-mat/0203046}, v.1: 3 March 2002, v.2: 21 October 2002,
   v.3: 5 November 2002}

\begin{abstract}
  We study the Heston model, where the stock price dynamics is
  governed by a geometrical (multiplicative) Brownian motion with
  stochastic variance.  We solve the corresponding Fokker-Planck
  equation exactly and, after integrating out the variance, find an
  analytic formula for the time-dependent probability distribution of
  stock price changes (returns).  The formula is in excellent
  agreement with the Dow-Jones index for the time lags from 1 to 250
  trading days.  For large returns, the distribution is exponential in
  log-returns with a time-dependent exponent, whereas for small
  returns it is Gaussian.  For time lags longer than the relaxation
  time of variance, the probability distribution can be expressed in a
  scaling form using a Bessel function.  The Dow-Jones data for
  1982--2001 follow the scaling function for seven orders of
  magnitude.
\end{abstract}

\maketitle

\section{Introduction}

Stochastic dynamics of stock prices is commonly described by a
geometric (multiplicative) Brownian motion, which gives a log-normal
probability distribution function (PDF) for stock price changes
(returns) \cite{Wilmott}.  However, numerous observations show that
the tails of the PDF decay slower than the log-normal distribution
predicts (the so-called ``fat-tails'' effect)
\cite{Bouchaud-book,Stanley-book,Voit}.  Particularly, much attention
was devoted to the power-law tails \cite{Mandelbrot,Stanley-returns}.
The geometric Brownian motion model has two parameters: the drift
$\mu$, which characterizes the average growth rate, and the volatility
$\sigma$, which characterizes the noisiness of the process.  There is
empirical evidence and a set of stylized facts indicating that
volatility, instead of being a constant parameter, is driven by a
mean-reverting stochastic process \cite{Engle,Papanicolaou}.  Various
mathematical models with stochastic volatility have been discussed in
literature
\cite{HullWhite,SteinStein,Heston,Baaquie,Bakshi,Duffie,Pan}.

In this paper, we study a particular stochastic volatility model,
called the Heston model \cite{Heston}, where the square of the
stock-price volatility, called the variance $v$, follows a random
process known in financial literature as the Cox-Ingersoll-Ross
process and in mathematical statistics as the Feller process
\cite{Papanicolaou,Feller}.  Using the Fourier and Laplace transforms
\cite{Feller,Duffie}, we solve the Fokker-Planck equation for this
model exactly and find the joint PDF of returns and variance as a
function of time, conditional on the initial value of variance.  While
returns are readily known from a financial time-series data, variance
is not given directly, so it acts as a hidden stochastic variable.
Thus, we integrate the joint PDF over variance and obtain the marginal
probability distribution function of returns \textit{un}conditional on
variance.  The latter PDF can be directly compared with financial
data.  We find an excellent agreement between our results and the
Dow-Jones data for the 20-years period of 1982--2001.  Using only four
fitting parameters, our equations very well reproduce the PDF of
returns for time lags between 1 and 250 trading days.  In contrast, in
ARCH, GARCH, EGARCH, TARCH, and similar models, the number of fitting
parameters can easily go to a few tens \cite{McMillan}.

Our result for the PDF of returns has the form of a one-dimensional
Fourier integral, which is easily calculated numerically or, in
certain asymptotical limits, analytically.  For large returns, we find
that the PDF is exponential in log-returns, which implies a power-law
distribution for returns, and we calculate the time dependence of the
corresponding exponents.  In the limit of long times, the PDF exhibits
scaling, i.e.\ it becomes a function of a single combination of return
and time, with the scaling function expressed in terms of a Bessel
function.  The Dow-Jones data follow the predicted scaling function
for seven orders of magnitude.

The original paper \cite{Heston} solved the problem of option pricing
for the Heston model.  Numerous subsequent studies
\cite{Bakshi,Duffie,Pan,options} compared option pricing derived
from this model and its extensions with the empirical data on option
pricing.  They found that the Heston model describes the empirical
option prices much better than the Black-Scholes theory, and
modifications of the Heston model, such as adding discontinuous jumps,
further improve the agreement.  However, these papers did not address
the fundamental question whether the stock market actually follows the
Heston stochastic process or not.  Obviously, if the answer is
negative, then using the Heston model for option pricing would not
make much sense.  The stock-market time series was studied in Ref.\ 
\cite{Pan} jointly with option prices, but the focus was just on
extracting the effective parameters of the Heston model.  In contrast,
we present a comprehensive comparison of the stock market returns
distribution with the predictions of the Heston model.  Using a single
set of four parameters, we fit the whole family of PDF curves for a
wide variety of time lags.  In order to keep the model as simple as
possible with the minimal number of fitting parameters, we use the
original Heston model and do not include later modifications proposed
in literature, such as jumps, multiple relaxation time, etc.
\cite{Bakshi,Duffie,Pan}.  Interestingly, the parameters of the
model that we find from our fits of the stock market data are of the
same order of magnitude as the parameters extracted from the fits of
option prices in Refs.\ \cite{Bakshi,Duffie,Pan}.

\section{The Model}
\label{sec:model}

We consider a stock, whose price $S_t$, as a function of time $t$,
obeys the stochastic differential equation of a geometric
(multiplicative) Brownian motion in the It\^{o} form
\cite{Wilmott,Gardiner}:
\begin{equation} \label{eqS}
  dS_t = \mu S_t\, dt + \sigma_t S_t\, dW_t^{(1)}.
\end{equation}
Here the subscript $t$ indicates time dependence, $\mu$ is the drift
parameter, $W_t^{(1)}$ is a standard random Wiener
process\footnote{The infinitesimal increments of the Wiener process
  $dW_t$ are normally-distributed (Gaussian) random variables with
  zero mean and the variance equal to $dt$.}, and $\sigma_t$ is the
time-dependent volatility.

Since any solution of (\ref{eqS}) depends only on $\sigma_t^2$, it is
convenient to introduce the new variable $v_t=\sigma_t^2$, which is
called the variance.  We assume that $v_t$ obeys the following
mean-reverting stochastic differential equation:
\begin{equation} \label{eqVar}
  dv_t = -\gamma(v_t - \theta)\,dt + \kappa\sqrt{v_t}\,dW_t^{(2)}.
\end{equation}
Here $\theta$ is the long-time mean of $v$, $\gamma$ is the rate of
relaxation to this mean, $W_t^{(2)}$ is a standard Wiener process, and
$\kappa$ is a parameter that we call the variance noise.  Eq.\ 
(\ref{eqVar}) is known in financial literature as the
Cox-Ingersoll-Ross (CIR) process and in mathematical statistics as the
Feller process \cite{Papanicolaou,Feller}.  Alternative equations for
$v_t$, with the last term in (\ref{eqVar}) replaced by
$\kappa\,dW_t^{(2)}$ or $\kappa v_t\,dW_t^{(2)}$, have been also
discussed in literature \cite{HullWhite}.  However, in our paper, we
study only the case given by Eq.\ (\ref{eqVar}).

We take the Wiener process appearing in (\ref{eqVar}) to be correlated
with the Wiener process in (\ref{eqS}):
\begin{equation} \label{rho}
   dW_t^{(2)} = \rho\,dW_t^{(1)} + \sqrt{1-\rho^2}\,dZ_t, 
\end{equation} 
where $Z_t$ is a Wiener process independent of $W_t^{(1)}$, and
$\rho\in[-1,1]$ is the correlation coefficient.  A negative
correlation ($\rho<0$) between $W_t^{(1)}$ and $W_t^{(2)}$ is known as
the leverage effect \cite[p.\ 41]{Papanicolaou}.

It is convenient to change the variable in (\ref{eqS}) from price
$S_t$ to log-return $r_t=\ln(S_t/S_0)$.  Using It\^{o}'s formula
\cite{Gardiner}, we obtain the equation satisfied by $r_t$:
\begin{equation}\label{eqR}
  dr_t = \left(\mu - \frac{v_t}{2}\right)dt +
  \sqrt{v_t}\,dW_t^{(1)}.
\end{equation}
The parameter $\mu$ can be eliminated from (\ref{eqR}) by changing the
variable to $x_t=r_t-\mu t$, which measures log-returns relative to
the growth rate $\mu$:
\begin{equation}\label{eqX}
   dx_t = - \frac{v_t}{2}\,dt + \sqrt{v_t}\,dW_t^{(1)}.
\end{equation}
Where it does not cause confusion with $r_t$, we use the term
``log-return'' also for the variable $x_t$.

Equations (\ref{eqX}) and (\ref{eqVar}) define a two-dimensional
stochastic process for the variables $x_t$ and $v_t$
\cite{Heston,Duffie}.  This process is characterized by the transition
probability $P_t(x,v\,|\,v_i)$ to have log-return $x$ and variance $v$
at time $t$ given the initial log-return $x=0$ and variance $v_i$ at
$t=0$.  Time evolution of $P_t(x,v\,|\,v_i)$ is governed by the
Fokker-Planck (or forward Kolmogorov) equation \cite{Gardiner}
\begin{eqnarray}\label{FP}
  && \frac{\partial}{\partial t}P = 
     \gamma\frac{\partial}{\partial v}\left[(v-\theta)P\right]
     + \frac12\frac{\partial}{\partial x}(vP)
\\ 
  && +\rho\kappa\frac{\partial^2}{\partial x\,\partial v}(vP)
     +\frac12\frac{\partial^2}{\partial x^2}(vP)  
     +\frac{\kappa^2}{2}\frac{\partial^2}{\partial v^2}(vP).  
\nonumber
\end{eqnarray}
The initial condition for (\ref{FP}) is a product of two delta
functions
\begin{equation} \label{initial}
        P_{t=0}(x,v\,|\,v_i)=\delta(x)\,\delta(v-v_i).
\end{equation}

\begin{figure}
\centerline{\epsfig{file=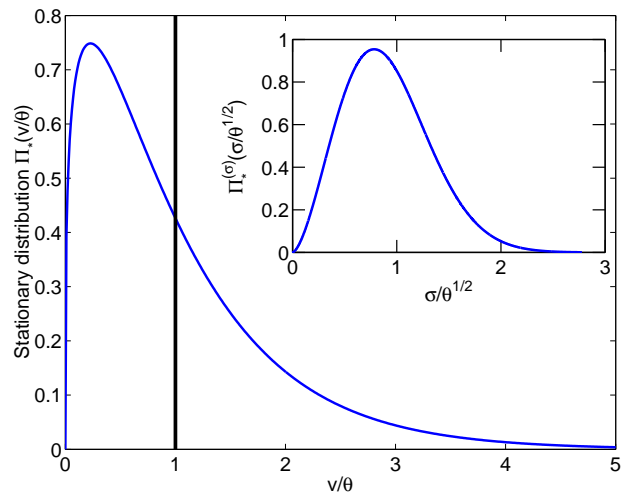,width=0.95\linewidth}}
\caption{The stationary probability distribution $\Pi_\ast(v)$ of
  variance $v$, given by Eq.\ (\ref{Pi_v}) and shown for $\alpha=1.3$
  from Table \ref{paramVal}. The vertical line indicates the average
  value of $v$.  Inset: The corresponding stationary probability
  distribution $\Pi_\ast^{(\sigma)}(v)$ of volatility $\sigma$ given
  by Eq.\ (\ref{Pi_s}). }
\label{fig:variance}
\end{figure}

The probability distribution of the variance itself, $\Pi_t(v)=\int
dx\,P_t(x,v)$, satisfies the equation
\begin{equation} \label{A1varFP}
  \frac{\partial}{\partial t}\Pi_t(v) = 
  \frac{\partial}{\partial v}\left[\gamma(v-\theta)\Pi_t(v)\right]
  +\frac{\kappa^2}{2}\frac{\partial^2}{\partial^2 v}\left[v\Pi_t(v)
  \right],
\end{equation}
which is obtained from (\ref{FP}) by integration over $x$.  Feller
\cite{Feller} has shown that this equation is well-defined on the
interval $v\in[0,+\infty)$ as long as $\theta>0$.  Eq.\
(\ref{A1varFP}) has the stationary solution
\begin{equation} \label{Pi_v}
   \Pi_\ast(v) = \frac{\alpha^\alpha}{\Gamma(\alpha)} \,
   \frac{v^{\alpha-1}}{\theta^\alpha} \,
   e^{-\alpha v/\theta}, \qquad 
   \alpha=\frac{2\gamma\theta}{\kappa^2},
\end{equation}
which is the Gamma distribution.  The parameter $\alpha$ in
(\ref{Pi_v}) is the ratio of the average variance $\theta$ to the
characteristic fluctuation of variance $\kappa^2/2\gamma$ during the
relaxation time $1/\gamma$.  When $\alpha\to\infty$,
$\Pi_\ast(v)\to\delta(v-\theta)$.  The corresponding stationary PDF of
volatility $\sigma$ is
\begin{equation} \label{Pi_s}
   \Pi_\ast^{(\sigma)}(\sigma) = \frac{2\alpha^\alpha}{\Gamma(\alpha)} 
   \frac{\sigma^{2\alpha-1}}{\theta^\alpha}
   e^{-\alpha\sigma^2/\theta}.
\end{equation}
Functions (\ref{Pi_v}) and (\ref{Pi_s}) are integrable as long as
$\alpha>0$.  The distributions $\Pi_\ast(v)$ and
$\Pi_\ast^{(\sigma)}(\sigma)$ are shown in Fig.\ \ref{fig:variance}
for the value $\alpha=1.3$ deduced from the fit of the Dow-Jones time
series and given in Table \ref{paramVal} in Sec.\ \ref{sec:data}.

\section{Solution of the Fokker-Planck equation}
\label{sec:FP}

Since $x$ appears in (\ref{FP}) only in the derivative operator
$\partial/\partial x$, it is convenient to take the Fourier transform
\begin{equation}\label{FT}
   P_t(x,v\,|\,v_i)=
   \int_{-\infty}^{+\infty}\frac{dp_x}{2\pi} \, 
   e^{ip_x x} \overline{P}_{t,p_x}(v\,|\,v_i).
\end{equation}
Inserting (\ref{FT}) into (\ref{FP}), we find
\begin{eqnarray} \label{FourSch}
   && \frac{\partial}{\partial t}\overline{P} = 
      \gamma\frac{\partial}{\partial v}
      \left[(v-\theta)\overline{P}\right]
\\
   && -\left[\frac{p_x^2 -ip_x}{2} v
   - i\rho\kappa p_x\frac{\partial}{\partial v} v 
   - \frac{\kappa^2}{2}\frac{\partial^2}{\partial v^2}
   v\right]\overline{P}.  \nonumber 
\end{eqnarray}
Eq.\ (\ref{FourSch}) is simpler than (\ref{FP}), because the number of
variables has been reduced to two, $v$ and $t$, whereas $p_x$ only
plays the role of a parameter.

Since Eq.~(\ref{FourSch}) is linear in $v$ and quadratic in
$\partial/\partial v$, it can be simplified by taking the Laplace
transform over $v$
\begin{equation} \label{FourPV}
   \widetilde{P}_{t,p_x}(p_v\,|\,v_i) =
   \int_{0}^{+\infty}\!\! dv \, e^{-p_v v}
   \overline{P}_{t,p_x}(v\,|\,v_i).
\end{equation}
The partial differential equation satisfied by
$\widetilde{P}_{t,p_x}(p_v\,|\,v_i)$ is of the first order
\begin{equation}\label{FourFour}
   \left[ \frac{\partial}{\partial t} + \left(
   \Gamma p_v+\frac{\kappa^2}{2}p_v^2 -\frac{p_x^2-ip_x}{2}
   \right)\frac{\partial}{\partial p_v}\right] \widetilde{P}
   = -\gamma\theta p_v \widetilde{P},
\end{equation}
where we introduced the notation 
\begin{equation} \label{Gamma}
    \Gamma = \gamma + i\rho\kappa p_x.
\end{equation}
Eq.~(\ref{FourFour}) has to be solved with the initial condition
\begin{equation} \label{v_i}
  \widetilde{P}_{t=0,p_x}(p_v\,|\,v_i)=\exp(-p_v v_i).
\end{equation}

The solution of (\ref{FourFour}) is given by the method of
characteristics \cite{CourantHilbert}:
\begin{equation} \label{Kpv}
   \widetilde{P}_{t,p_x}(p_v\,|\,v_i) = \exp\left(-\tilde{p}_v(0)v_i
   -\gamma\theta\int_{0}^{t} d\tau\,\tilde{p}_v(\tau)\right),
\end{equation}
where the function $\tilde{p}_v(\tau)$ is the solution of the
characteristic (ordinary) differential equation
\begin{equation}\label{charEq}
   \frac{d\tilde{p}_v(\tau)}{d\tau} = \Gamma \tilde{p}_v(\tau) 
   +\frac{\kappa^2}{2}\tilde{p}_v^2(\tau)  
   -\frac{p_x^2 - ip_x}{2}
\end{equation}
with the boundary condition $\tilde{p}_v(t)=p_v$ specified at
$\tau=t$.  The differential equation (\ref{charEq}) is of the Riccati
type with constant coefficients \cite{Bender}, and its solution is
\begin{equation} \label{pvSol}
  \tilde{p}_v(\tau) =  
  \frac{2\Omega}{\kappa^2}\frac{1}{\zeta e^{\Omega(t-\tau)} - 1} 
  -\frac{\Gamma-\Omega}{\kappa^2},
\end{equation}
where we introduced the frequency
\begin{equation} \label{eqOmega}
   \Omega=\sqrt{\Gamma^2 + \kappa^2(p_x^2-ip_x)}.  
\end{equation}
and the coefficient 
\begin{equation} \label{zeta}
   \zeta = 1 + \frac{2\Omega}{\kappa^2 p_v +(\Gamma-\Omega)}.
\end{equation}
Substituting (\ref{pvSol}) into (\ref{Kpv}), we find
\begin{eqnarray} \label{solution}
   && \widetilde{P}_{t,p_x}(p_v\,|\,v_i) 
\\
   && =\exp\left\{-\tilde{p}_v(0)v_i
   + \frac{\gamma\theta(\Gamma-\Omega)t}{\kappa^2}
   - \frac{2\gamma\theta}{\kappa^2}\ln
   \frac{\zeta - e^{-\Omega t}}{\zeta - 1}\right\}.
\nonumber
\end{eqnarray}

\section{Averaging over variance}
\label{sec:averaging}

Normally we are interested only in log-returns $x$ and do not care
about variance $v$.  Moreover, whereas log-returns are directly known
from financial data, variance is a hidden stochastic variable that has
to be estimated.  Inevitably, such an estimation is done with some
degree of uncertainty, which precludes a clear-cut direct comparison
between $P_t(x,v\,|\,v_i)$ and financial data.  Thus we introduce the
reduced probability distribution
\begin{equation} \label{P}
   P_t(x\,|\,v_i)=\int\limits_{0}^{+\infty}\!\!dv\,P_t(x,v\,|\,v_i)
   =\int\frac{dp_x}{2\pi}e^{i p_x x}\widetilde{P}_{t,p_x}(0\,|\,v_i),
\end{equation}
where the hidden variable $v$ is integrated out, so $p_v=0$.
Substituting $\zeta$ from (\ref{zeta}) with $p_v=0$ into
(\ref{solution}), we find
\begin{eqnarray} \label{finalex}
   && P_t(x\,|\,v_i) = \int_{-\infty}^{+\infty} 
   \frac{dp_x}{2\pi}\, e^{i p_x x 
   - v_i\frac{p_x^2 - ip_x}{\Gamma + \Omega\coth{(\Omega t/2)}} }
\nonumber \\
   && \times\, e^{- \frac{2\gamma\theta}{\kappa^2}\ln\left(
   \cosh\frac{\Omega t}{2} +\frac{\Gamma}{\Omega}\sinh\frac{\Omega t}{2}
   \right) + \frac{\gamma\Gamma \theta t}{\kappa^2}}.
\end{eqnarray}

To check the validity of (\ref{finalex}), let us consider the limiting
case $\kappa=0$.  In this case, the stochastic term in (\ref{eqVar})
is absent, so the time evolution of variance is deterministic:
\begin{equation} \label{detV}
  v_t= \theta + (v_i - \theta)e^{-\gamma t}.
\end{equation}
Then process (\ref{eqX}) gives a Gaussian distribution for $x$,
\begin{equation}\label{K0}
   P_t^{(\kappa=0)}(x\,|\,v_i)=\frac{1}{\sqrt{2\pi t \overline{v}_t}}
   \exp\left(-\frac{(x+\overline{v}_t t/2)^2}{2\overline{v}_t t}\right),
\end{equation}
with the time-averaged variance $\overline{v}_t =
\frac{1}{t}\int_{0}^{t}d\tau\,v_\tau$.  On the other hand, by taking
the limit $\kappa\rightarrow0$ and integrating over $p_x$ in
(\ref{finalex}), we reproduce the same expression (\ref{K0}).

Eq.\ (\ref{finalex}) cannot be directly compared with financial
time-series data, because it depends on the unknown initial variance
$v_i$.  In order to resolve this problem, we assume that $v_i$ has the
stationary probability distribution $\Pi_\ast(v_i)$, which is given by
(\ref{Pi_v}).  Thus we introduce the probability distribution function
$P_t(x)$ by averaging (\ref{finalex}) over $v_i$ with the weight
$\Pi_\ast(v_i)$:
\begin{equation} \label{dv_i}
  P_t(x)= \int_0^\infty \!\!dv_i\,\Pi_\ast(v_i)\,P_t(x\,|\,v_i).
\end{equation}
The integral over $v_i$ is similar to the one of the Gamma function
and can be taken explicitly.  The final result is the Fourier integral
\begin{equation} \label{Pfinal}
   P_t(x) = \frac{1}{2\pi}\int_{-\infty}^{+\infty} \!\!dp_x\,
   e^{ip_x x + F_t(p_x)}
\end{equation}
with
\begin{eqnarray}
  && F_t(p_x)=\frac{\gamma\theta}{\kappa^2}\, \Gamma t
\label{phaseF} \\
  && {} - \frac{2\gamma\theta}{\kappa^2}
  \ln\left[\cosh\frac{\Omega t}{2} + 
  \frac{\Omega^2 -\Gamma^2 + 2\gamma\Gamma}{2\gamma\Omega}
  \sinh\frac{\Omega t}{2}\right].
\nonumber
\end{eqnarray}

\begin{figure}
\centerline{\epsfig{file=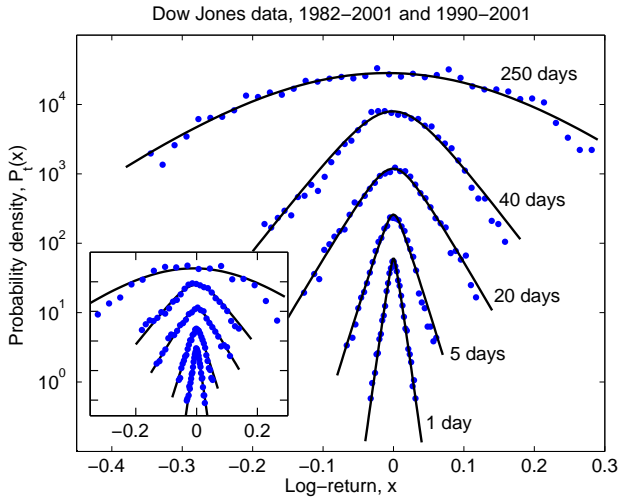,width=0.95\linewidth}}
\caption{ Probability distribution $P_t(x)$ of log-return $x$ for
  different time lags $t$.  Points: The 1982--2001 Dow-Jones data for
  $t=1$, 5, 20, 40, and 250 trading days.  Solid lines: Fit of the
  data with Eqs.\ (\ref{Pfinal}) and (\ref{phaseF}).  For clarity, the
  data points and the curves for successive $t$ are shifted up by the
  factor of 10 each.  Inset: The 1990--2001 Dow-Jones data points
  compared with the same theoretical curves. }
\label{fig:data}
\end{figure}

The variable $p_x$ enters (\ref{phaseF}) via the variables $\Gamma$
from (\ref{Gamma}) and $\Omega$ from (\ref{eqOmega}).  It is easy to
check that $P_t(x)$ is real, because Re$F$ is an even function of
$p_x$ and Im$F$ is an odd one.  One can also check that
$F_t(p_x=0)=0$, which implies that $P_t(x)$ is correctly normalized at
all times: $\int dx\,P_t(x)=1$.  The simplified version of Eq.\ 
(\ref{phaseF}) for the case $\rho=0$ is given in Appendix
\ref{sec:rho=0}.

Eqs.\ (\ref{Pfinal}) and (\ref{phaseF}) for the probability
distribution $P_t(x)$ of log-return $x$ at time $t$ are the central
analytical result of the paper.  The integral in (\ref{Pfinal}) can be
calculated numerically or, in certain regimes discussed in Secs.\ 
\ref{t>>1}, \ref{x>>1}, and \ref{t<<1}, analytically.  In Fig.\ 
\ref{fig:data}, the calculated function $P_t(x)$, shown by solid
lines, is compared with the Dow-Jones data, shown by dots.  (Technical
details of the data analysis are discussed in Sec.\ \ref{sec:data}.)
Fig.\ \ref{fig:data} demonstrates that, with a fixed set of the
parameters $\gamma$, $\theta$, $\kappa$, $\mu$, and $\rho$, Eqs.\ 
(\ref{Pfinal}) and (\ref{phaseF}) very well reproduce the distribution
of log-returns $x$ of the Dow-Jones index for \textit{all} times $t$.
In the log-linear scale of Fig.\ \ref{fig:data}, the tails of $\ln
P_t(x)$ vs.\ $x$ are straight lines, which means that that tails of
$P_t(x)$ are exponential in $x$.  For short times $t$, the
distribution is narrow, and the slopes of the tails are nearly
vertical.  As the time $t$ progresses, the distribution broadens and
flattens.

\section{Asymptotic behavior for long time $t$}
\label{t>>1}

Eq.\ (\ref{eqVar}) implies that variance reverts to the equilibrium
value $\theta$ within the characteristic relaxation time $1/\gamma$.
In this section, we consider the asymptotic limit where time $t$ is
much longer than the relaxation time: $\gamma t\gg2$.  According to
(\ref{Gamma}) and (\ref{eqOmega}), this condition also implies that
$\Omega t\gg2$.  Then Eq.\ (\ref{phaseF}) reduces to
\begin{equation} \label{F_t>>1}
   F_t(p_x)\approx\frac{\gamma\theta t}{\kappa^2}(\Gamma-\Omega).
\end{equation}

Let us change of the variable of integration in (\ref{Pfinal}) to
\begin{equation} \label{tilde_p_x}
   p_x=\frac{\omega_0}{\kappa\sqrt{1-\rho^2}}\,\tilde p_x+ip_0,
\end{equation}
   where
\begin{equation} \label{p_0}
   p_0=\frac{\kappa-2\rho\gamma}{2\kappa(1-\rho^2)},\quad
   \omega_0=\sqrt{\gamma^2+\kappa^2(1-\rho^2)p_0^2}.
\end{equation}
Substituting (\ref{tilde_p_x}) into (\ref{Gamma}), (\ref{eqOmega}), and
(\ref{F_t>>1}), we transform (\ref{Pfinal}) to the following form
\begin{equation}\label{AppAll}
   P_t(x) = \frac{\omega_0e^{-p_0x+\Lambda t}}
   {\pi\kappa\sqrt{1-\rho^2}}
   \int_0^\infty\!\!d\tilde p_x\, \cos(A\tilde p_x) 
   e^{-B\sqrt{1+\tilde p_x^2}}, 
\end{equation}
where 
\begin{equation}
   A=\frac{\omega_0}{\kappa\sqrt{1-\rho^2}}\left(x 
   + \rho\frac{\gamma\theta t}{\kappa}\right),\quad
   B=\frac{\gamma\theta\omega_0 t}{\kappa^2},
\end{equation}
and
\begin{equation} \label{Lambda}
   \Lambda=\frac{\gamma\theta}{2\kappa^2}
   \frac{2\gamma-\rho\kappa}{1-\rho^2}.
\end{equation}
According to formula 3.914 from \cite{GR}, the integral in
(\ref{AppAll}) is equal to $B K_1(\sqrt{A^2+B^2})/\sqrt{A^2+B^2}$,
where $K_1$ is the first-order modified Bessel function.

\begin{figure}
\centerline{\epsfig{file=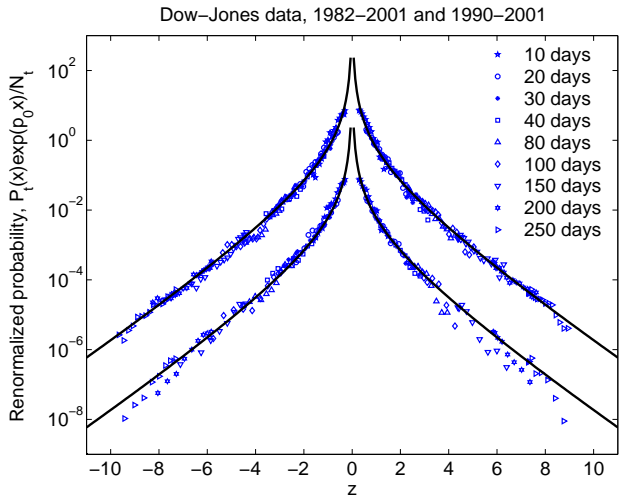,width=0.95\linewidth}}
\caption{Renormalized probability density $P_t(x)e^{p_0x}/N_t$ plotted
  as a function of the scaling argument $z$ given by (\ref{K1arg}).
  Solid lines: The scaling function $P_{\ast}(z)=K_1(z)/z$ from
  (\ref{Pbess}), where $K_1$ is the first-order modified Bessel
  function.  Upper and lower sets of points: The 1982--2001 and
  1990--2001 Dow-Jones data for different time lags $t$.  For clarity,
  the lower data set and the curve are shifted by the factor of
  $10^{-2}$.  }
\label{fig:Bessel}
\end{figure}

Thus, Eq.\ (\ref{Pfinal}) in the limit $\gamma t\gg2$ can be
represented in the scaling form
\begin{equation} \label{Pbess}
   P_t(x)=N_t\,e^{-p_0x}P_{\ast}(z), \quad P_{\ast}(z)=K_1(z)/z,
\end{equation}
where the argument $z=\sqrt{A^2+B^2}$ is
\begin{equation} \label{K1arg}
   z=\frac{\omega_0}{\kappa}
   \sqrt{\frac{(x+\rho\gamma\theta t/\kappa)^2}{1-\rho^2} + 
   \left(\frac{\gamma\theta t}{\kappa}\right)^2},
\end{equation}
and the time-dependent normalization factor $N_t$ is
\begin{equation} \label{N}
   N_t=\frac{\omega_0^2\gamma\theta t}{\pi\kappa^3\sqrt{1-\rho^2}}
   \,e^{\Lambda t},
\end{equation}
Eq.\ (\ref{Pbess}) demonstrates that, up to the factors $N_t$ and
$e^{-p_0x}$, the dependence of $P_t(x)$ on the two arguments $x$ and
$t$ is given by the function $P_{\ast}(z)$ of the single scaling
argument $z$ in (\ref{K1arg}).  Thus, when plotted as a function of
$z$, the data for different $x$ and $t$ should collapse on the single
universal curve $P_{\ast}(z)$. This is beautifully illustrated by
Fig.\ \ref{fig:Bessel}, where the Dow-Jones data for different time
lags $t$ follows the curve $P_{\ast}(z)$ for seven orders of
magnitude.

In the limit $z\gg1$, we can use the asymptotic expression \cite{GR}
$K_1(z)\approx e^{-z}\sqrt{\pi/2z}$ in (\ref{Pbess}) and take the
logarithm of $P$.  Keeping only the leading term proportional to $z$
and omitting the subleading term proportional to $\ln z$, we find that
$\ln P_t(x)$ has the hyperbolic distribution \cite[p.\ 
14]{Bouchaud-book}
\begin{equation} \label{P32}
   \ln\frac{P_t(x)}{N_t}\approx -p_0x-z \quad {\rm for} \quad z\gg1.
\end{equation}
Let us examine Eq.\ (\ref{P32}) for large and small $|x|$.

In the first case $|x|\gg\gamma\theta t/\kappa$, Eq.\ (\ref{K1arg})
gives $z\approx\omega_0|x|/\kappa\sqrt{1-\rho^2}$, so Eq.\ (\ref{P32})
becomes
\begin{equation} \label{PlargeX}
   \ln\frac{P_t(x)}{N_t}
   \approx-p_0x-\frac{\omega_0}{\kappa\sqrt{1-\rho^2}}|x|.
\end{equation}
Thus, the PDF $P_t(x)$ has the exponential tails (\ref{PlargeX}) for
large log-returns $|x|$.  Notice that, in the considered limit $\gamma
t\gg2$, the slopes $d\ln P/dx$ of the exponential tails
(\ref{PlargeX}) do not depend on time $t$.  Because of $p_0$, the
slopes (\ref{PlargeX}) for positive and negative $x$ are not equal,
thus the distribution $P_t(x)$ is not symmetric with respect to
positive and negative price changes.  According to (\ref{p_0}), this
asymmetry is enhanced by a negative correlation $\rho<0$ between stock
price and variance.

\begin{figure}
\centerline{\epsfig{file=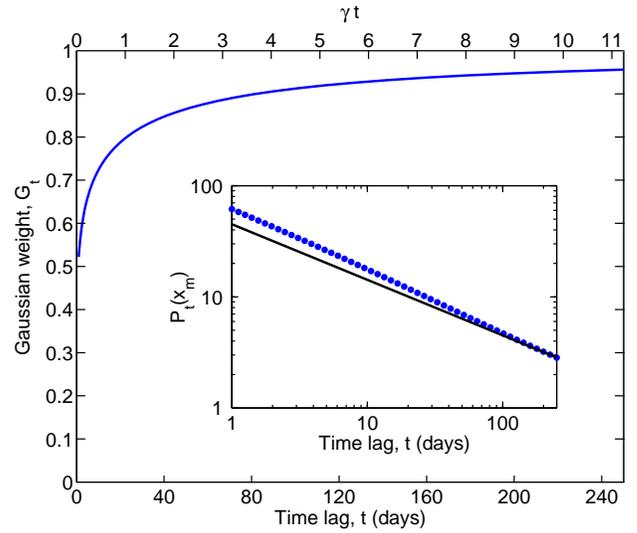,width=0.95\linewidth}}
\caption{The fraction $G_t$ of the total probability contained in the
   Gaussian part of $P_t(x)$ vs.\ time lag $t$.  Inset: Time
   dependence of the probability density at maximum $P_t(x_m)$
   (points), compared with the Gaussian $t^{-1/2}$ behavior (solid
   line).}
\label{fig:Gauss}
\end{figure}

In the second case $|x+\rho\gamma\theta t/\kappa| \ll\gamma\theta
t/\kappa$, by Taylor-expanding $z$ in (\ref{K1arg}) near its minimum
in $x$ and substituting the result into (\ref{P32}), we get
\begin{equation} \label{Gauss}
    \ln\frac{P_t(x)}{N'_t}\approx -p_0x -
    \frac{\omega_0(x+\rho\gamma\theta t/\kappa)^2}
    {2(1-\rho^2)\gamma\theta t},
\end{equation}
where $N'_t=N_t\exp(-\omega_0\gamma\theta t/\kappa^2)$.  Thus, for
small log-returns $|x|$, the PDF $P_t(x)$ is Gaussian with the width
increasing linearly in time.  The maximum of $P_t(x)$ in (\ref{Gauss})
is achieved at
\begin{equation} \label{x_m}
   x_{m}(t)=-\frac{\gamma\theta t}{2\omega_0}\left(
   1+2\,\frac{\rho(\omega_0-\gamma)}{\kappa}\right).
\end{equation}
Eq.\ (\ref{x_m}) gives the most probable log-return $x_m(t)$ at time
$t$, and the coefficient in front of $t$ constitutes a correction to
the average growth rate $\mu$, so that the actual growth rate is
$\bar\mu=\mu+d x_m/dt$.

As Fig.\ \ref{fig:data} illustrates, $\ln P_t(x)$ is indeed linear in
$x$ for large $|x|$ and quadratic for small $|x|$, in agreement with
(\ref{PlargeX}) and (\ref{Gauss}).  As time progresses, the
distribution, which has the scaling form (\ref{Pbess}) and
(\ref{K1arg}), broadens.  Thus, the fraction $G_t$ of the total
probability contained in the parabolic (Gaussian) portion of the curve
increases, as illustrated in Fig.\ \ref{fig:Gauss}.  (The procedure of
calculating $G_t$ is explained in Appendix \ref{sec:Gauss}.)  Fig.\ 
\ref{fig:Gauss} shows that, at sufficiently long times, the total
probability contained in the non-Gaussian tails becomes negligible,
which is known in literature \cite{Bouchaud-book}.  The inset in Fig.\ 
\ref{fig:Gauss} illustrates that the time dependence of the
probability density at maximum, $P_t(x_m)$, is close to $t^{-1/2}$,
which is characteristic for a Gaussian evolution.

\begin{figure}
\centerline{\epsfig{file=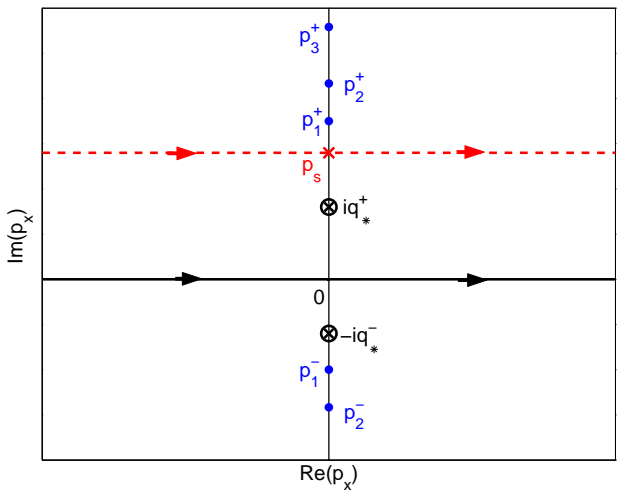,width=0.95\linewidth}}
\caption{ Complex plane of $p_x$. Dots: The singularities of
  $F_t(p_x)$. Circled crosses: The accumulation points $\pm iq_*^\pm$
  of the singularities in the limit $\gamma t\gg 2$.  Cross: The
  saddle point $p_s$, which is located in the upper half-plane for
  $x>0$.  Dashed line: The contour of integration displaced from the
  real axis in order to pass through the saddle point $p_s$.}
\label{fig:complex} 
\end{figure}

\section{Asymptotic behavior for large log-return $x$}
\label{x>>1}

In the complex plane of $p_x$, function $F(p_x)$ becomes singular at
the points $p_x$ where the argument of the logarithm in (\ref{phaseF})
vanishes.  These points are located on the imaginary axis of $p_x$ and
are shown by dots in Fig.\ \ref{fig:complex}.  The singularity closest
to the real axis is located on the positive (negative) imaginary axis
at the point $p_1^+$ $(p_1^-)$.  Because the argument of the logarithm
in (\ref{phaseF}) vanishes at these two points, we can approximate
$F(p_x)$ by the dominant, singular term:
$F(p_x)\approx-(2\gamma\theta/\kappa^2)\ln(p_x-p_1^\pm)$.

\begin{figure}
\centerline{\epsfig{file=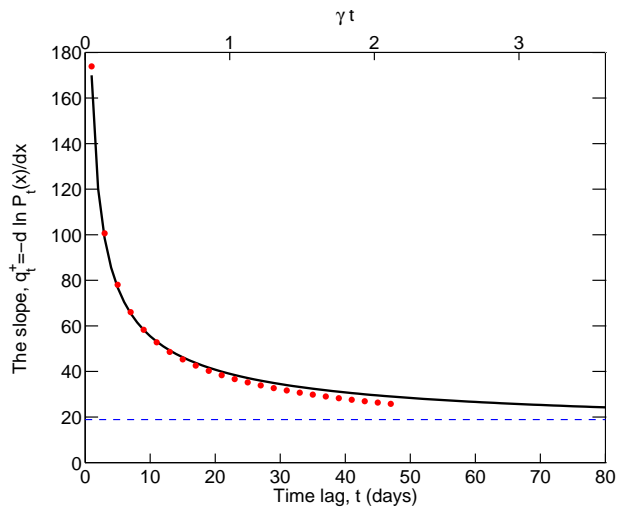,width=0.95\linewidth}}
\caption{ Solid line: The slope $q_t^+=-d\ln P/dx$ of the exponential
   tail for $x>0$ as a function of time.  Points: The asymptotic
   approximation (\ref{p1pm}) for the slope in the limit $\gamma
   t\ll2$.  Dashed line: The saturation value $q_*^+$ for $\gamma
   t\gg2$, Eq.\ (\ref{q_*}).}
\label{fig:slope} 
\end{figure}

For large $|x|$, the integrand of (\ref{Pfinal}) oscillates very fast
as a function of $p_x$.  Thus, we can evaluate the integral using the
method of stationary phase \cite{Bender} by shifting the contour of
integration so that is passes through a saddle point of the argument
$ip_xx+F(p_x)$ of the exponent in (\ref{Pfinal}).  The saddle point
position $p_s$, shown in Fig.\ \ref{fig:complex} by the cross, is
determined by the equation
\begin{equation} \label{saddle}
   ix=-\left.\frac{dF(p_x)}{dp_x}\right|_{p_x=p_s}
   \approx\frac{2\gamma\theta}{\kappa^2}\times
   \left\{\begin{array}{ll}
      \frac{1}{p_s-p_1^+},& x>0, \\
      \frac{1}{p_s-p_1^-},& x<0.
   \end{array}\right.
\end{equation}
For a large $|x|$, such that $|xp_1^\pm|\gg2\gamma\theta/\kappa^2$, the
saddle point $p_s$ is very close to the singularity point: $p_s\approx
p_1^+$ for $x>0$ and $p_s\approx p_1^-$ for $x<0$.  Then the
asymptotic expression for the probability distribution is
\begin{equation} \label{P_x>>1}
   P_t(x)\sim
   \left\{\begin{array}{ll}
      e^{-xq_t^+},& x>0, \\
      e^{ xq_t^-},& x<0,
   \end{array}\right.
\end{equation}
where $q_t^\pm=\mp ip_1^\pm(t)$ are real and positive.  Eq.\
(\ref{P_x>>1}) shows that, for all times $t$, the tails of the
probability distribution $P_t(x)$ for large $|x|$ are exponential.
The slopes of the exponential tails, $q^\pm=\mp\,d\ln P/dx$, are
determined by the positions $p_1^{\pm}$ of the singularities closest
to the real axis.
 
These positions $p_1^{\pm}(t)$ and, thus, the slopes $q_t^\pm$ depend
on time $t$.  For times much shorter than the relaxation time ($\gamma
t\ll2$), the singularities lie far away from the real axis.  As time
increases, the singularities move along the imaginary axis toward the
real axis.  Finally, for times much longer than the relaxation time
($\gamma t\gg2$), the singularities approach limiting points:
$p_1^{\pm}\to\pm iq_*^\pm$, which are shown in Fig.\ \ref{fig:complex}
by circled crosses.  Thus, as illustrated in Fig.\ \ref{fig:slope},
the slopes $q_t^\pm$ monotonously decrease in time and saturate at
long times:
\begin{equation} \label{q_*}
   q_t^\pm\to q_*^\pm=\frac{\omega_0}{\kappa\sqrt{1-\rho^2}} \pm p_0
   \quad {\rm for} \quad \gamma t\gg2.
\end{equation}
The slopes (\ref{q_*}) are in agreement with Eq.\ (\ref{PlargeX})
valid for $\gamma t\gg2$.  The time dependence $q_t^{\pm}$ at short
times can be also found analytically:
\begin{equation} \label{p1pm}
   q_t^{\pm}\approx\frac{2}{\kappa}\sqrt{\frac{\gamma}{t}}
   \quad {\rm for} \quad \gamma t\ll2.
\end{equation}
The dotted curve in Fig.\ \ref{fig:slope} shows that Eq.\ (\ref{p1pm})
works very well for short times $t$, where the slope diverges at
$t\to0$.

\section{Asymptotic behavior for short time $t$}
\label{t<<1}

For a short time $t$, we expand the equations of Sec.\ \ref{sec:FP} to
the first order in $t$ and set $p_v=0$.  The last term in Eq.\ 
(\ref{solution}) cancels the penultimate term, and Eq.\ (\ref{charEq})
gives $\tilde{p}_v(0)=t(p_x^2-ip_x)/2$.  Substituting this formula
into (\ref{solution}) and taking the integral (\ref{P}) over $p_x$, we
find
\begin{equation} \label{Ptv_i}
   P_t(x\,|\,v_i)=\frac{1}{\sqrt{2\pi v_i t}}\,
   e^{-\frac{(x+v_it/2)^2}{2v_it}}.
\end{equation}
Eq.\ (\ref{Ptv_i}) shows that, for a short $t$, the probability
distribution of $x$ evolves in a Gaussian manner with the initial
variance $v_i$, because variance has no time to change.

Substituting (\ref{Ptv_i}) and (\ref{Pi_v}) into (\ref{dv_i}), we find
\begin{equation} \label{Pcanon}
   P_t(x)=\frac{\alpha^{\alpha}}{\Gamma(\alpha)}\frac{e^{-x/2}}
   {\sqrt{2\pi \theta t}}\int_0^\infty\!\! d\tilde v_i\, 
   \tilde v_i^{C-1} e^{-A \tilde v_i - B/\tilde v_i}
\end{equation}
where $\tilde v_i=v_i/\theta$, $A=\alpha+\theta t/8$, $B=x^2/2\theta t$,
and $C=\alpha-1/2$.  According to formula 3.471.9 from \cite{GR}, the
integral in (\ref{Pcanon}) is $2(B/A)^{C/2}K_C(2\sqrt{AB})$ for
Re$A>0$ and Re$B>0$, where $K_C$ is the modified Bessel function of
the order $C$.  Taking into account that $A\approx \alpha$ (because
$t\ll16\gamma/\kappa^2$ for short $t$), we obtain the final expression
\begin{equation} \label{PshortF}
   P_t(x)=\frac{2^{1-\alpha}e^{-x/2}}{\Gamma(\alpha)}\,
   \sqrt{\frac{\alpha}{\pi \theta t}}\,
   y^{\alpha-1/2}
   K_{\alpha-1/2}(y),
\end{equation}
where we introduced the scaling variable 
\begin{equation} \label{y}
   y=\sqrt{\frac{2\alpha x^2}{\theta t}}
   =\frac{2\sqrt{\gamma}}{\kappa}\,\frac{|x|}{\sqrt{t}}
\end{equation}

In the limit $y\gg1$, using the formula $K_\nu(y)\approx
e^{-y}\sqrt{\pi/2y}$ in (\ref{PshortF}), we find
\begin{equation} \label{y>>1}
   P_t(x)\approx\frac{2^{1/2-\alpha}}{\Gamma(\alpha)}\,
   \sqrt{\frac{\alpha}{\theta t}}\,
   y^{\alpha-1}e^{-y}.
\end{equation}
Eqs.\ (\ref{y}) and (\ref{y>>1}) show that the tails of the
distribution are exponential in $x$, and the slopes $d\ln P/dx$ are in
agreement with Eq.\ (\ref{p1pm}).

In the opposite limit $y\ll1$, the small argument expansion of the
Bessel function can be found from the following equations
\cite{Stegun}:
\begin{equation} \label{defK}
   K_\nu(y)=\frac{\pi}{2}\frac{I_{-\nu}(y)-I_\nu(y)}{\sin(\nu\pi)},
   \quad \frac{\pi}{\sin(\pi\nu)}=\Gamma(\nu)\Gamma(1-\nu),
\end{equation}
and 
\begin{equation} \label{seriesI}
   I_\nu(y)\approx\left(\frac y2\right)^\nu\sum_{k=0}^{\infty} 
   \frac{(y^2/4)^k}{k!\,\Gamma(\nu+k+1)}.
\end{equation}
Substituting (\ref{seriesI}) into (\ref{defK}), we find in the case
$1/2\le\alpha<3/2$
\begin{eqnarray}
   K_{\alpha-1/2}(y)&\approx&
   \frac{\Gamma(\alpha-1/2)}{2} \left(\frac y2\right)^{-\alpha+1/2}
\nonumber \\
   && {}+\frac{\Gamma(-\alpha+1/2)}{2} 
   \left(\frac y2\right)^{\alpha-1/2}.
\label{K}
\end{eqnarray}
Substituting (\ref{K}) into (\ref{PshortF}), we obtain
\begin{equation} \label{y<<1}
   P_t(x)\approx\frac{\Gamma(\alpha-1/2)}{\Gamma(\alpha)}\,
   \sqrt{\frac{\alpha}{2\pi\theta t}}
   \left[1-\lambda\left(\frac{y}{2}\right)
   ^{2\alpha-1}\right],
\end{equation}
where we introduced the coefficient
\begin{equation}
   \lambda=\frac{|\Gamma(-\alpha+1/2)|}{\Gamma(\alpha-1/2)}.
\end{equation}
Eq.\ (\ref{y<<1}) can be written in the form
\begin{equation} \label{lnP,y<<1}
   \ln P_t(x) - \ln P_t(0) \approx
   -\lambda\left(\frac{y}{2}\right)^{2\alpha-1}.
\end{equation}
We see that $\ln P_t(x)$ approaches $x=0$ as a power of $x$ lower than
2 (for $1/2\le\alpha<3/2$).  The slope $d\ln P/dx$ at $x\to0$ is zero
for $\alpha>1$ and infinite for $\alpha<1$.

\section{Comparison with the Dow-Jones time series}
\label{sec:data}

To test the model against financial data, we downloaded daily closing
values of the Dow-Jones industrial index for the period of 20 years
from 1 January 1982 to 31 December 2001 from the Web site of Yahoo
\cite{Yahoo}. The data set contains 5049 points, which form the time
series $\{S_\tau\}$, where the integer time variable $\tau$ is the
trading day number.  We do not filter the data for short days, such as
those before holidays.

Given $\{S_\tau\}$, we use the following procedure to extract the
probability density $P_t^{(DJ)}(r)$ of log-return $r$ for a given time
lag $t$.  For the fixed $t$, we calculate the set of log-returns
$\{r_\tau=\ln S_{\tau+t}/S_\tau\}$ for all possible times $\tau$.
Then we partition the $r$-axis into equally spaced bins of the width
$\Delta r$ and count the number of log-returns $r_\tau$ belonging to
each bin.  In this process, we omit the bins with occupation numbers
less than five, because we consider such a small statistics
unreliable.  Only less than 1\% of the entire data set is omitted in
this procedure.  Dividing the occupation number of each bin by $\Delta
r$ and by the total occupation number of all bins, we obtain the
probability density $P_t^{(DJ)}(r)$ for a given time lag $t$.  To find
$P_t^{(DJ)}(x)$, we replace $r\to x+\mu t$.

Assuming that the system is ergodic, so that ensemble averaging is
equivalent to time averaging, we compare $P_t^{(DJ)}(x)$ extracted
from the time-series data and $P_t(x)$ calculated in previous
sections, which describes ensemble distribution.  In the language of
mathematical statistics, we compare our theoretically derived
population distribution with the sample distribution extracted from
the time series data.  We determine parameters of the model by
minimizing the mean-square deviation $\sum_{x,t}|\ln P_t^{(DJ)}(x)-\ln
P_t(x)|^2$, where the sum is taken over all available $x$ and $t=1$,
5, 20, 40, and 250 days.  These values of $t$ are selected because
they represent different regimes: $\gamma t\ll1$ for $t=1$ and 5 days,
$\gamma t\approx1$ for $t=20$ days, and $\gamma t\gg1$ for $t=40$ and
250 days.  As Figs.\ \ref{fig:data} and \ref{fig:Bessel} illustrate,
our expression (\ref{Pfinal}) and (\ref{phaseF}) for the probability
density $P_t(x)$ agrees with the data very well, not only for the
selected five values of time $t$, but for the whole time interval from
1 to 250 trading days.  However, we cannot extend this comparison to
$t$ longer than 250 days, which is approximately 1/20 of the entire
range of the data set, because we cannot reliably extract
$P_t^{(DJ)}(x)$ from the data when $t$ is too long.

The values obtained for the four fitting parameters ($\gamma$,
$\theta$, $\kappa$, $\mu$) are given in Table \ref{paramVal}.  We find
that our fits are not very sensitive to the value of $\rho$, so we
cannot reliably determine it.  Thus, we use $\rho=0$ for simplicity,
which gives a good fit of the data.  On the other hand, a nonzero
value of $\rho$ was found in \cite{Masoliver} by fitting the leverage
correlation function introduced in \cite{Bouchaud-PRL} and in
\cite{Bakshi,Duffie,Pan} by fitting the option prices.

All four parameters ($\gamma$, $\theta$, $\kappa$, $\mu$) shown in
Table \ref{paramVal} have the dimensionality of 1/time.  The first
line of the Table gives their values in the units of 1/day, as
originally determined in our fit.  The second line shows the
annualized values of the parameters in the units of 1/year, where we
utilize the average number of 252.5 trading days per calendar year to
make the conversion.  The relaxation time of variance is equal to
$1/\gamma=22.2$ trading days = 4.4 weeks $\approx$ 1 month, where we
took into account that 1 week = 5 trading days.  Thus, we find that
variance has a rather long relaxation time, of the order of one month,
which is in agreement with the conclusion of Ref.\ \cite{Masoliver}.

\begin{table}  
\caption{\label{paramVal} 
  Parameters of the Heston model obtained from the fit of the 
  Dow-Jones data using $\rho=0$ for the correlation coefficient.  
  We also find $1/\gamma=22.2$ trading days for the relaxation time 
  of variance, $\alpha=2\gamma\theta/\kappa^2=1.3$ for the parameter 
  in the variance distribution function (\ref{Pi_v}), and
  $x_0=\kappa/\gamma=5.4\%$ for the characteristic scale 
  (\ref{tildes}) of $x$. }
\begin{tabular}{c|cccc}
\hline
Units & $\gamma$ & $\theta$  &  $\kappa$ & $\mu$ \\
\hline
1/day  &  $4.50\times10^{-2}$   &  $8.62\times10^{-5}$  &  
       $2.45\times10^{-3}$      &  $5.67\times10^{-4}$ \\
\hline
1/year  &  11.35   &  0.022   & 0.618   &  0.143 \\
\hline
\end{tabular}
\end{table}

\begin{figure}
\centerline{\epsfig{file=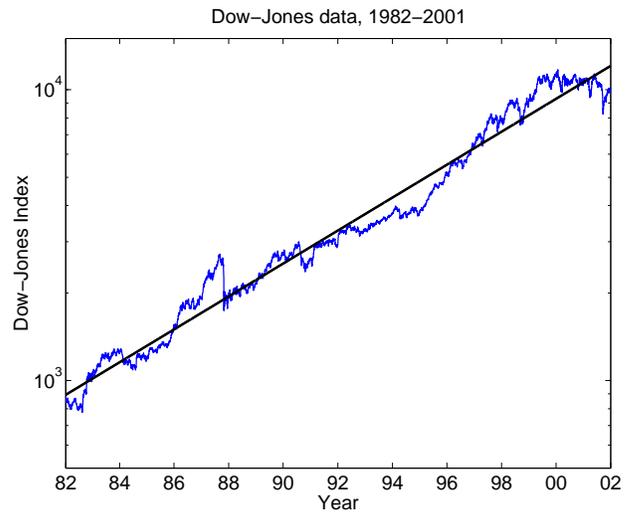,width=0.95\linewidth}}
\caption{Time dependence of the Dow-Jones index shown in the
  log-linear scale.  The straight line represents the average
  exponential growth in time.}
\label{fig:Dow} 
\end{figure}

The effective growth rate of stock prices is determined by the
coordinate $r_m(t)$ where the probability density $P_t(r_m)$ is
maximal.  Using the relation $r_m=x_m+\mu t$ and Eq.\ (\ref{x_m}), we
find that the actual growth rate is
$\bar\mu=\mu-\gamma\theta/2\omega_0\approx\mu-\theta/2=13$\% per year.
[Here we took into account that $\omega_0\approx\gamma$, because
$\gamma\gg\kappa/2$ in Eq.\ (\ref{p_0}).]  This number coincides with
the average growth rate of the Dow-Jones index obtained by a simple
fit of the time series $\{S_\tau\}$ with an exponential function of
$\tau$, as shown in Fig.\ \ref{fig:Dow}.  The effective stock growth
rate $\bar{\mu}$ is comparable with the average stock volatility after
one year $\sigma=\sqrt{\theta}=14.7\%$.  Moreover, as Fig.\
\ref{fig:variance} shows, the distribution of variance is broad, and
the variation of variance is comparable to its average value $\theta$.
Thus, even though the average growth rate of stock index is positive,
there is a substantial probability $\int_{-\infty}^0
dr\,P_t(r)=17.7$\% to have negative return for $t=1$ year.

According to (\ref{q_*}), the asymmetry between the slopes of
exponential tails for positive and negative $x$ is given by the
parameter $p_0$, which is equal to 1/2 when $\rho=0$ (see also the
discussion of Eq.\ (\ref{Pfinal'}) in Appendix \ref{sec:rho=0}).  The
origin of this asymmetry can be traced back to the transformation from
(\ref{eqS}) to (\ref{eqR}) using It\^{o}'s formula.  It produces the
term $0.5v_t\,dt$ in the r.h.s.\ of (\ref{eqR}), which then generates
the second term in the r.h.s.\ of (\ref{FP}).  The latter term is the
only source of asymmetry in $x$ of $P_t(x)$ when $\rho=0$.  However,
in practice, the asymmetry of the slopes $p_0=1/2$ is quite small
(about 2.7\%) compared with the average slope
$q_*^\pm\approx\omega_0/\kappa\approx1/x_0=18.4$.

By  fitting the Dow-Jones data to   our formula, we implicitly assumed
that the parameters  of  the stochastic  process  ($\gamma$, $\theta$,
$\kappa$, $\mu$) do not change in time.   While this assumption may be
reasonable for a limited time interval, the parameters generally could
change in  time.  The time interval of  our fit,  1982--2001, includes
the crash of 1987, so one might expect  that the parameters of the fit
would  change   if we  use   a different   interval.  To  verify  this
conjecture, in  Figs.\  \ref{fig:data} and  \ref{fig:Bessel},  we also
compare  the data points  for  the time  interval 1990--2001 with  the
theoretical curves produced using  the same values for  the parameters
as shown in Table \ref{paramVal}.  Although  the empirical data points
in  the tails  for long  time lags decrease  somewhat faster  than the
theory predicts, the overall agreement  is quite reasonable.  We  find
that changing the  values of the  fitting parameters does  not visibly
improve the agreement.  Thus, we  conclude that the parameters of  the
Heston   stochastic  process are essentially   the  same for 1980s and
1990s.  Apparently, the  crash of 1987  produced little  effect on the
probability distribution of  returns, because the stock market quickly
resumed its overall growth.  On the other hand, the study \cite{Silva}
indicates that the data for 2000s do not follow our theoretical curves
with  the same fitting parameters.   The main difference appears to be
in the average  growth rate $\mu$,  which became negative in 2000s, as
opposed  to +13\%  per  year in  1980s and  1990s.  Unfortunately, the
statistics for 2000s  is  limited, because we   have  only few  years.
Nevertheless, it does seem to indicate  that in 2000s the stock market
switched to a different regime compared with 1980s and 1990s.

\section{Discussion and conclusions}

We derived an analytical solution for the PDF $P_t(x)$ of log-returns
$x$ as a function of time $t$ for the Heston model of a geometrical
Brownian motion with stochastic variance.  The final result has the
form of a one-dimensional Fourier integral (\ref{Pfinal}) and
(\ref{phaseF}).  (In the case $\rho=0$, the equations have the simpler
form presented in Appendix \ref{sec:rho=0}.)  Our result agrees very
well with the Dow-Jones data, as shown in Fig.\ \ref{fig:data}.
Comparing the theory and the data, we determine the four (non-zero)
fitting parameters of the model, particularly the variance relaxation
time $1/\gamma=22.2$ days.  For time longer than $1/\gamma$, our
theory predicts the scaling behavior (\ref{Pbess}) and (\ref{K1arg}),
which the Dow-Jones data indeed exhibits over seven orders of
magnitude, as shown in Fig.\ \ref{fig:Bessel}.  The scaling function
$P_{\ast}(z)=K_1(z)/z$ is expressed in terms of the first-order
modified Bessel function $K_1$.  Previous estimates in literature of
the relaxation time of volatility using various indirect indicators
range from 1.5 days \cite[p.\ 80]{Papanicolaou} to 73 days for the
half-life of the Dow-Jones index \cite{Engle}.  Since we have a good
fit of the entire family of PDFs for time lags from 1 to 250 trading
days, we believe that our estimate, 22.2 days, is more reliable.  A
close value of 19.6 days was found in Ref.\ \cite{Masoliver}.

An alternative point of view in literature is that the time evolution
of volatility is not characterized by a single relaxation rate.  As
shown in Appendix \ref{sec:correlation}, the variance correlation
function $C_t^{(v)}$ (\ref{Cv-t}) in the Heston model has a simple
exponential decay in time.  However, the analysis of financial data
\cite[p.\ 70]{Bouchaud-book} indicates that the correlation function
has a power-law dependence or superposition of two (or more)
exponentials with the relaxation times of less than one day and more
than few tens of days. (Large amount of noise in the data makes it
difficult to give a precise statement.)  Ref.\ \cite{Bacry} argues
that volatility relaxation is multifractal and has no characteristic
time.  However, one should keep in mind that the total range
(\ref{Cv-0infty}) of variation of $C_t^{(v)}$ is only about 77\% of
its saturation value, not many orders of magnitude.  As Figs.\ 2 of
Refs.\ \cite{Bouchaud-EL} and \cite{Bacry} shows, the main drop of
$C_t^{(v)}$ takes place within a reasonably well-defined and
relatively short time, whereas residual relaxation is stretched over a
very long time.  In this situation, a simple exponential time
dependence, while not exact, may account for the main part of
relaxation and give a reasonable approximation for the purposes of our
study.  Alternatively, it is possible to generalize the Heston model
by incorporating more than one relaxation times \cite{Duffie}.

As Fig.\ \ref{fig:data} shows, the probability distribution $P_t(x)$
is exponential in $x$ for large $|x|$, where it is characterized by
the time-dependent slopes $d\ln P/dx$.  The theoretical analysis
presented in Sec.\ \ref{x>>1} shows that the slopes are determined by
the singularities of the function $F_t(p_x)$ from (\ref{phaseF}) in
the complex plane of $p_x$ that are closest to the real axis.  The
calculated time dependence of the slopes $d\ln P/dx$, shown in Fig.\
\ref{fig:slope}, agrees with the data very well, which further
supports our statement that $1/\gamma=22.2$ days.  Exponential tails
in the probability distribution of stock log-returns have been noticed
in literature before \cite[p.\ 61]{Bouchaud-book}, \cite{Miranda},
however time dependence of the slopes has not been recognized and
analyzed theoretically.  As shown in Fig.\ \ref{fig:data}, our
equations give the parabolic dependence of $\ln P_t(x)$ on $x$ for
small $x$ and linear dependence for large $x$, in agreement with the
data.  Qualitatively similar results were found in Ref.\ \cite{Serva}
for a different model with stochastic volatility and in agreement with
the NYSE index daily data.  It suggests that the linear and parabolic
behavior is a generic feature of the models with stochastic
volatility.  In Ref.\ \cite{Stanley-returns}, the power-law dependence
on $x$ of the tails of $P_t(x)$ was emphasized.  However, the data for
S\&P 500 were analyzed in Ref.\ \cite{Stanley-returns} only for short
time lags $t$, typically shorter than one day.  On the other hand, our
data analysis is performed for the time lags longer than one day, so
the results cannot be directly compared.

Deriving $P_t(x)$ in Sec.\ \ref{sec:averaging}, we assumed that
variance $v$ has the stationary gamma-distribution $\Pi_*(v)$
(\ref{Pi_v}).  This assumption should be compared with the data.
There were numerous attempts in literature to reconstruct the
probability distribution of volatility from the time-series data
\cite{Stanley-volatility,Mantegna-volatility}.  Generally, these
papers agree that the central part of the distribution is well
described by a log-normal distribution, but opinions vary on the
fitting of the tails.  Particularly, Ref.\ \cite{Mantegna-volatility}
performed a fit with an alternative probability distribution of
volatility described in Ref.\ \cite[p.\ 88]{Bouchaud-book}.
Unfortunately, none of these papers attempted to fit the data using
Eq.\ (\ref{Pi_s}), so we do not have a quantitative comparison.
Taking into account that we only need the integral (\ref{dv_i}), the
exact shape of $\Pi_*(v)$ may be not so important, and Eq.\
(\ref{Pi_v}) may give a reasonably good approximation for our
purposes, even if it does not fit the tails very precisely.

Although we tested our model for the Dow-Jones index, there is nothing
specific in the model which indicates that it applies only to stock
market data.  It would be interesting to see how the model performs
when applied to other time series, for example, the foreign exchange
data \cite{Peinke}, which also seem to exhibit exponential tails.  The
study \cite{Silva} indicates that our $P_t(x)$ also works very well
for the S\&P 500 and Nasdaq indices for 1980s and 1990s.  However, in
the 2000s the average growth rate $\mu$ of the stock market changed to
a negative value, which complicates separation of fluctuations from
the overall trend.

\begin{acknowledgments}
  We thank J.-P.~Bouchaud and R.~E.~Prange for detailed discussions of
  the paper and numerous valuable comments, and A.~T.~Zheleznyak and
  A.~C.~Silva for careful reading of the manuscript and finding minor
  errors.
\end{acknowledgments}

\appendix

\section{The case \boldmath $\rho=0$}
\label{sec:rho=0}

As explained in Sec.\ \ref{sec:data}, we fit the data using $\rho=0$
for simplicity.  In this case, by shifting the variable of integration
in (\ref{Pfinal}) $p_x\to p_x+i/2$, we find
\begin{equation} \label{Pfinal'}
   P_t(x) = e^{-x/2}\int_{-\infty}^{+\infty} \frac{dp_x}{2\pi} \,
   e^{ip_x x + F_t(p_x)},
\end{equation}
where $\alpha=2\gamma\theta/\kappa^2$,
\begin{equation} \label{phaseF'}
   F_t(p_x)=\frac{\alpha\gamma t}{2}  
   - \alpha\ln\left[\cosh\frac{\Omega t}{2} + 
   \frac{\Omega^2+\gamma^2}{2\gamma\Omega}
   \sinh\frac{\Omega t}{2}\right],
\end{equation}
and
\begin{equation} \label{eqOmega'}
   \Omega=\sqrt{\gamma^2 + \kappa^2(p_x^2+1/4)}
   \approx\gamma\sqrt{1+p_x^2(\kappa^2/\gamma^2)}.
\end{equation}
Now the function $F_t(p_x)$ is real and symmetric:
$F_t(p_x)=F_t(-p_x)$.  Thus, the integral in (\ref{Pfinal'}) is a
symmetric function of $x$, and the only source of asymmetry of
$P_t(x)$ in $x$ is the exponential prefactor in (\ref{Pfinal'}), as
discussed at the end of Sec.\ \ref{sec:data}.

In the second equation (\ref{eqOmega'}), we took into account that,
according to values shown in Table \ref{paramVal},
$\kappa^2/4\gamma^2\ll1$.  Introducing the dimensionless variables
\begin{equation} \label{tildes}
  \tilde t=\gamma t,\quad \tilde x=x/x_0,
  \quad \tilde p_x=p_xx_0  ,\quad x_0=\kappa/\gamma,
\end{equation}
Eqs.\ (\ref{Pfinal'}), (\ref{phaseF'}), and (\ref{eqOmega'}) can be
rewritten as follows:
\begin{equation} \label{Pfinal'-}
   P_t(x) = \frac{e^{-x/2}}{x_0}\int_{-\infty}^{+\infty} 
   \frac{d\tilde p_x}{2\pi} \,
   e^{i\tilde p_x \tilde x + F_{\tilde t}(\tilde p_x)},
\end{equation}
where $\tilde\Omega=\sqrt{1+\tilde p_x^2}$ and
\begin{equation} \label{phaseF'-}
   F_{\tilde t}(\tilde p_x)=\frac{\alpha\tilde t}{2}  
   - \alpha\ln\left[\cosh\frac{\tilde\Omega\tilde t}{2} + 
   \frac{\tilde\Omega^2+1}{2\tilde\Omega}
   \sinh\frac{\tilde\Omega\tilde t}{2}\right].
\end{equation}
It is clear from (\ref{tildes}), (\ref{Pfinal'-}), and
(\ref{phaseF'-}) that the parameter $\alpha$ determines the shape of
the function $P_t(x)$, whereas $1/\gamma$ and $x_0$ set the scales of
$t$ and $x$.  

In the limit $\tilde t\gg2$, the scaling function (\ref{Pbess}) for
$\rho=0$ can be written as
\begin{equation}
   P_t(x)=N_t\,e^{-x/2}K_1(z)/z,\quad
   z=\sqrt{\tilde x^2 + \bar t^2},
\label{Pbess-}
\end{equation}
where $\bar t=\alpha\tilde t/2=t\theta/x_0^2$ and $N_t=\bar te^{\bar
  t}/\pi x_0$.  Notice that Eq.\ (\ref{Pbess-}) has only two fitting
parameters, $x_0$ and $\theta$, whereas the general formula
(\ref{Pfinal'-}) and (\ref{phaseF'-}) has three fitting parameters.
As follows from (\ref{P32}), for $\tilde x\gg\bar t$ and $\tilde
x\gg1$, $P_t(x)\propto\exp(-|x|/x_0)$, so $1/x_0$ is the slope of the
exponential tails in $x$.

\section{Gaussian weight}
\label{sec:Gauss}

Let us expand the integral in (\ref{Pfinal'}) for small $x$:
\begin{equation} \label{expand}
   P_t(x)\approx e^{-x/2}\left(\mu_0-\frac12\mu_2 x^2\right),
\end{equation}
where the coefficients are the first and the second moments of
$\exp[F_t(p_x)]$
\begin{equation} \label{mu}
   \mu_0(t)=\int\limits_{-\infty}^{+\infty}
   \frac{dp_x}{2\pi}\,e^{F_t(p_x)},
   \quad 
   \mu_2(t)=\int\limits_{-\infty}^{+\infty}
   \frac{dp_x}{2\pi}\,p_x^2e^{F_t(p_x)}.
\end{equation}
On the other hand, we know that $P_t(x)$ is Gaussian for small $x$.
So, we can write
\begin{equation} \label{approxGauss}
   P_t(x)\approx \mu_0\, e^{-x/2}e^{-\mu_2 x^2/2\mu_0},
\end{equation}
with the same coefficients as in (\ref{expand}).  If we ignore the
existence of fat tails and extrapolate (\ref{approxGauss}) to
$x\in(-\infty,\infty)$, the total probability contained in such a
Gaussian extrapolation will be
\begin{equation} \label{W}
   G_t=\int\limits_{-\infty}^{+\infty}dx\,\mu_0\,e^{-x/2-\mu_2
   x^2/2\mu_0}=\sqrt{\frac{2\pi\mu_0^3}{\mu_2}}\,
   e^{\mu_0/8\mu_2}.
\end{equation}
Obviously, $G_t<1$, because the integral (\ref{W}) does not take into
account the probability contained in the fat tails.  Thus, the
difference $1-G_t$ can be taken as a measure of how much the actual
distribution $P_t(x)$ deviates from a Gaussian function.

We calculated the moments (\ref{mu}) numerically for the function $F$
given by (\ref{phaseF'}), then determined the Gaussian weight $G_t$
from (\ref{W}) and plotted it in Fig.\ \ref{fig:Gauss} as a function
of time.  For $t\to\infty$, $G_t\to1$, i.e.\ $P_t(x)$ becomes Gaussian
for very long time lags, which is known in literature
\cite{Bouchaud-book}.  In the opposite limit $t\to0$, $F_t(p_x)$
becomes a very broad function of $p_x$, so we cannot calculate the
moments $\mu_0$ and $\mu_2$ numerically.  The singular limit $t\to0$
is studied analytically in Sec.\ \ref{t<<1}.

\section{Correlation function of variance}
\label{sec:correlation}

The correlation function of variance is defined as
\begin{equation} \label{Cv}
  C_t^{(v)}=\langle v_{t+\tau}v_\tau\rangle
  =\int_0^\infty dv_i \int_0^\infty dv\,
  v\,\Pi_t(v\,|\,v_i)\,v_i\,\Pi_*(v_i).
\end{equation}
It depends only on the relative time $t$ and does not depend on the
initial time $\tau$.  The averaging $\langle\ldots\rangle$ is
performed over the ensemble probability distribution, as written in
(\ref{Cv}), or over the initial time $\tau$ for time-series data.
Eq.\ (\ref{Cv}) has the same structure as in the influence-functional
formalism of Feynman and Vernon \cite{Feynman}, where
$\Pi_t(v\,|\,v_i)$ represents the conditional probability propagator
from the initial value $v_i$ to the final value $v$ over the time $t$,
and $\Pi_*(v_i)$ represents the stationary, equilibrium probability
distribution of $v_i$.

Using Eqs.\ (\ref{Cv}) and (\ref{Pi_v}), it is easy to find the
limiting values of $C_t^{(v)}$:
\begin{eqnarray}
  && C_\infty^{(v)}=\langle v\rangle^2=\theta^2,
\label{Cv-0infty}  \\
  && C_0^{(v)}=\langle v^2\rangle=\theta^2
  \left(1+\frac{1}{\alpha}\right)=\theta^2(1+0.77),
\nonumber
\end{eqnarray}
where we used the numerical value from Table \ref{paramVal}.

Differentiating Eq.\ (\ref{Cv}) with respect to $t$ and using Eq.\
(\ref{A1varFP}), we find that $C_t^{(v)}$ satisfies the following
differential equation:
\begin{equation} \label{Cv-dt}
  \frac{dC_t^{(v)}}{dt}=-\gamma(C_t^{(v)}-\theta^2).
\end{equation}
Thus, $C_t^{(v)}$ changes in time exponentially with the relaxation
rate $\gamma$:
\begin{equation} \label{Cv-t}
  C_t^{(v)}=\theta^2
  \left(1+\frac{e^{-\gamma t}}{\alpha}\right).
\end{equation}


\end{document}